\documentclass{article}
\usepackage[utf8]{inputenc}
\usepackage[T1]{fontenc}
\usepackage{lmodern}
\usepackage{amssymb}
\usepackage{amsmath}
\usepackage{graphicx}
\usepackage{float}
\usepackage{booktabs}
\usepackage{geometry}
\usepackage{xcolor}
\usepackage{tikz}
\usetikzlibrary{arrows.meta,positioning,calc,fit,decorations.pathreplacing}

\usepackage[colorlinks=true,
            linkcolor=blue!60!black,
            citecolor=green!50!black,
            urlcolor=blue!60!black,
            breaklinks=true,
            bookmarksnumbered=true]{hyperref}
\newcommand{\ket}[1]{\left|#1\right\rangle}
\newcommand{\bra}[1]{\left\langle#1\right|}

\title{Fermionic 2+1D SU(2) quantum link models: an ED/DMRG reliability study}
\author{Daming Li\thanks{lidaming@sjtu.edu.cn} \\ School of Mathematical Sciences,
Shanghai Jiao Tong University, Shanghai, 200240, China}
\date{}

\begin{document}
\maketitle

\begin{abstract}
Non-Abelian lattice gauge theory with dynamical fermions is a central target for
tensor-network and quantum-simulation methods, but controlled reliability
studies on small systems are scarce. We benchmark two complementary
Hilbert-space approaches for a 2+1D SU(2) quantum link model with one staggered
fermion flavour on small two-dimensional lattices: sparse exact diagonalization (ED) and
matrix-product-state density-matrix renormalization group (DMRG). The link variables use the five-state $\{5\}$
truncation, and
fermions are represented by two-colour four-state sites. We implement ED with
explicit cross-site Jordan--Wigner strings and DMRG with total-fermion-number
conservation and a matrix product operator (MPO) built from local terms. We validate the link algebra,
Gauss-law commutation, Hermiticity, the pure-gauge limit, and the MPO against
ED component by component. The $2\times2$ methods agree to $10^{-13}$ and the
$2\times3$--$2\times5$ ladders are controlled at or below the $10^{-5}$ level. The $3\times3$ geometry illustrates the role of colour neutrality: at
half filling the odd fermion number makes $G^2=3/4$ the exact lower bound, while
one hole or one doublon restores neutrality and lowers the residual to
$\sim2\times10^{-5}$ at $\chi=2048$. An independent exact-Gauss fPEPS provides a
reference for the even-filling $3\times2$ sector.
The results define a reproducible
reliability envelope for non-Abelian gauge-fermion benchmarks and show where
larger-scale simulations require larger bond dimensions or different
tensor-network geometries.
\end{abstract}

\section{Introduction}\label{sec:intro}

Lattice gauge theory provides a nonperturbative definition of QCD
\cite{Wilson1974,Kogut1979,Rothe2005}. Hamiltonian formulations are natural for
tensor networks \cite{Orus2014,Cirac2021}, neural quantum states
\cite{Carleo2017,Torlai2018}, and quantum circuits \cite{Preskill2018,Banuls2020}.
The main obstacles are the gauge constraint, the non-Abelian link Hilbert space,
and, for dynamical fermions, the Jordan--Wigner (JW) strings and the sign
problem.

The 2+1D SU(2) quantum link model (QLM) is a useful testbed. Its link variables
can be truncated to a small representation \cite{Ludwig2026}, while the model
retains non-Abelian gauge invariance. Adding a single staggered fermion flavour
introduces the essential gauge-fermion coupling without the full complexity of
QCD. The $2\times2$ lattice is small enough for ED, while the $2\times2$--$2\times5$
ladders and the small $3\times2$/$3\times3$ geometries are accessible to DMRG.

Tensor-network studies of Hamiltonian lattice gauge theory began in 1+1
dimensions, where the Schwinger model is the standard testbed. Matrix-product
states have been used for spectral calculations, string breaking, continuum
limits, and topological sectors
\cite{Banuls2013,Buyens2014,Buyens2016,Buyens2017,Funcke2020}. For non-Abelian
theories, K\"uhn \emph{et al.} studied SU(2) string breaking \cite{Kuhn2015},
Ba\~nuls \emph{et al.} introduced an efficient physical basis for 1+1D SU(2)
spectral calculations \cite{Banuls2017}, and Silvi \emph{et al.} simulated 1+1D
SU(3) at finite density \cite{Silvi2019}.

In 2+1 dimensions, tensor-network and quantum-link approaches have been
developed for U(1) and SU(2) gauge theories
\cite{Rico2014,Pichler2016,Felser2020,Cataldi2024}. Cataldi \emph{et al.}
\cite{Cataldi2024} recently simulated (2+1)D SU(2) Yang--Mills theory at finite
density with tensor networks; this is the closest methodological precedent to
our work. Tensor renormalization group methods provide a complementary
Euclidean formulation \cite{Meurice2022,Fukuma2021}.

Gauge-invariant tensor-network constructions, including entanglement
renormalization with gauge symmetry and symmetric tensor networks, have been
developed to enforce the Gauss law exactly
\cite{TagliacozzoVidal2011,Silvi2014,Zache2023}. Our work differs in scope:
rather than targeting a phase diagram or a continuum limit, we fix a small set
of parameter points and ask which energies and plaquettes are numerically
controlled. We provide an independent ED reference, a component-by-component
MPO check, and a diagnosis of the half-filled $3\times3$ geometry, where an odd
fermion number makes the Gauss law unsatisfiable.

The term benchmark here refers to numerical reliability, not to hardware
characterization: randomized benchmarking and gate-set tomography
\cite{Knill2008,BlumeKohout2017} estimate gate fidelities and SPAM on quantum
devices, whereas this work validates a classical tensor-network computation
against truncation, the Gauss penalty, and local-minima errors.

In this paper we ask a deliberately narrow question: \emph{which results for
this model are numerically reliable, and how do we know?} We do not attempt a
continuum extrapolation or a phase diagram. Instead we build a validation
protocol and apply it at a small number of fixed parameter points. Our
contributions are:
\begin{enumerate}
\item a five-state SU(2) QLM implementation with a staggered fermion and an
explicit JW convention;
\item a sparse ED reference and an independent N-conserving MPO/DMRG
implementation;
\item a component-by-component MPO/ED comparison, including the hopping term
that is most sensitive to JW signs;
\item a reliability envelope for $2\times2$--$2\times5$ ladders, including
Gauss residuals, fermion-number conservation, and initial-state checks;
\item the filling-parity obstruction on the half-filled $3\times3$ geometry
and an even-$N_f$ $3\times2$ control validated against an exact-Gauss reference.
\end{enumerate}

The paper is organized as follows. Section~\ref{sec:model} fixes the lattice,
the five-state link, the staggered fermion and its Jordan--Wigner convention,
the Hamiltonian, and the Gauss law. Section~\ref{sec:methods} describes the
sparse ED reference and the $N$-conserving MPO/DMRG pipeline, including the
lattice-to-chain mapping and the convergence rules. Section~\ref{sec:validation}
presents the exact identities and the component-by-component benchmarks.
Section~\ref{sec:results} reports the $2\times3$--$2\times5$ ladders and the
$3\times3$/$3\times2$ geometries. Section~\ref{sec:discussion} discusses the
reliability envelope and the two failure modes (symmetry obstruction and
penalty-limited under-convergence), and states the limitations. Appendices~\ref{app:params}--
\ref{app:scaling} collect the numerical parameters and per-point diagnostics and
the MPS scaling data; Appendix~\ref{app:worked} works out the $2\times2$ case
explicitly.

\section{Model}\label{sec:model}

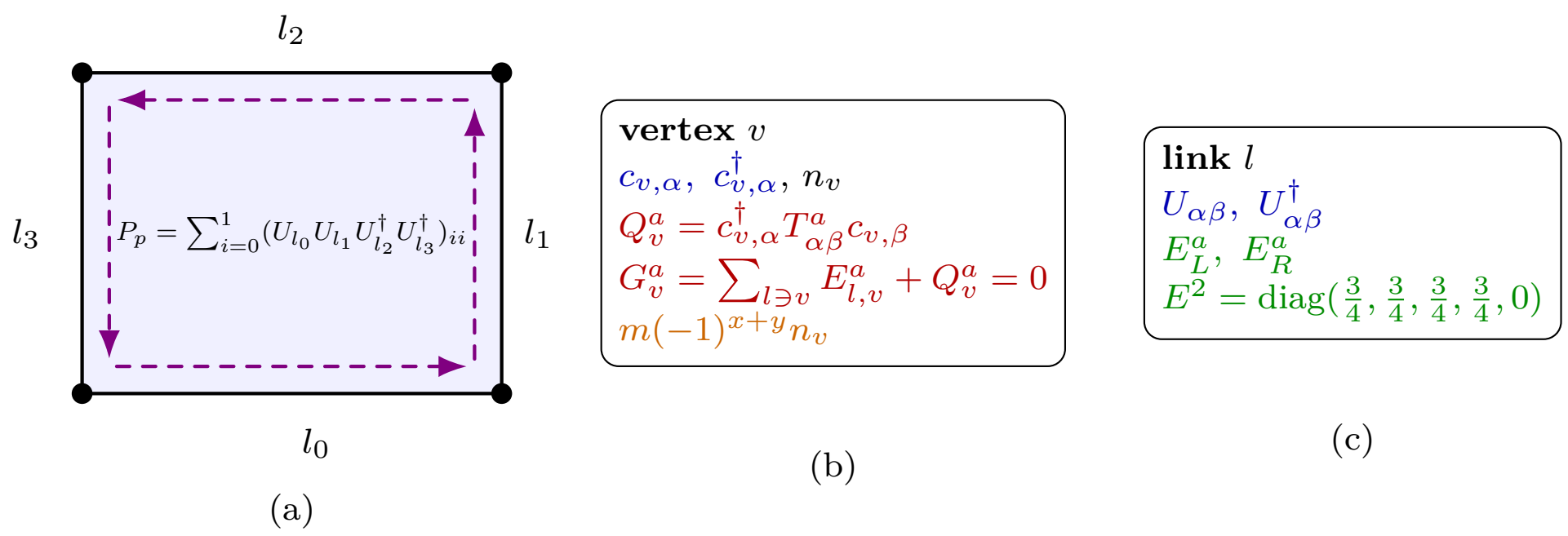
\begin{figure}[H]
\centering
\begin{tikzpicture}[font=\footnotesize, >=Latex, line cap=round, line join=round]

\fill[blue!6] (0,0) rectangle (3.4,2.6);
\draw[thick] (0,0)--(3.4,0);
\draw[thick] (3.4,0)--(3.4,2.6);
\draw[thick] (3.4,2.6)--(0,2.6);
\draw[thick] (0,2.6)--(0,0);

\draw[->, dashed, violet, thick] (0.28,0.22)--(3.12,0.22);
\draw[->, dashed, violet, thick] (3.18,0.28)--(3.18,2.32);
\draw[->, dashed, violet, thick] (3.12,2.38)--(0.28,2.38);
\draw[->, dashed, violet, thick] (0.22,2.32)--(0.22,0.28);

\node[fill=white, inner sep=1pt] at (1.9,-0.4) {$l_0$};
\node[fill=white, inner sep=1pt] at (3.7,1.3) {$l_1$};
\node[fill=white, inner sep=1pt] at (1.7,2.95) {$l_2$};
\node[fill=white, inner sep=1pt] at (-0.45,1.3) {$l_3$};

\foreach \p in {(0,0),(3.4,0),(0,2.6),(3.4,2.6)}
  \fill[black] \p circle (2.4pt);

\node at (1.7,1.3)
  {\resizebox{2.85cm}{!}{$P_p=\sum_{i=0}^{1}(U_{l_0}U_{l_1}U^\dagger_{l_2}U^\dagger_{l_3})_{ii}$}};

\node[draw, rounded corners, fill=white, align=left, anchor=west,
      inner sep=4pt] (vbox) at (4.2,1.3) {%
\textbf{vertex} $v$\\[1pt]
\textcolor{blue!70!black}{$c_{v,\alpha},\ c^\dagger_{v,\alpha}$},
$n_v$\\
\textcolor{red!70!black}{$Q_v^a=c^\dagger_{v,\alpha}T^a_{\alpha\beta}
  c_{v,\beta}$}\\
\textcolor{red!70!black}{$G_v^a=\sum_{l\ni v}E_{l,v}^a+Q_v^a=0$}\\
\textcolor{orange!80!black}{$m(-1)^{x+y}n_v$}};

\node[draw, rounded corners, fill=white, align=left, anchor=west,
      inner sep=4pt] (lbox) at (8.6,1.3) {%
\textbf{link} $l$\\[1pt]
\textcolor{blue!70!black}{$U_{\alpha\beta},\ U^\dagger_{\alpha\beta}$}\\
\textcolor{green!55!black}{$E_L^a,\ E_R^a$}\\
\textcolor{green!55!black}{$E^2=\mathrm{diag}(\tfrac34,\tfrac34,\tfrac34,
  \tfrac34,0)$}};

\node at (1.7,-0.95) {(a)};
\node[below=0.55cm of vbox] {(b)};
\node[below=0.55cm of lbox] {(c)};

\end{tikzpicture}
\caption{Operator content of the model. (a) A single plaquette of the lattice,
with the links $l_0,\dots,l_3$ ordered counter-clockwise, the plaquette operator
$P_p=\sum_{i=0}^{1}(U_{l_0}U_{l_1}U^\dagger_{l_2}U^\dagger_{l_3})_{ii}$, the standard
staggered phases $\eta_x=1$ (horizontal) and $\eta_y=(-1)^x$ (vertical), and the
four sites. (b) Operators acting on a vertex $v$: the two-colour fermion modes
$c_{v,\alpha}$, $c^\dagger_{v,\alpha}$, the number $n_v$, the colour charge
$Q_v^a$, the Gauss constraint $G_v^a=0$, and the mass term. (c) Operators acting
on a link $l$: $U_{\alpha\beta}$, $U^\dagger_{\alpha\beta}$, $E_L^a$, $E_R^a$,
and $E^2$. The ladder geometries used in the text are obtained by repeating the
vertical direction.}
\label{fig:model}
\end{figure}

This section fixes the lattice, the five-state link with its electric and link
operators, the staggered fermion and its Jordan--Wigner convention, the
Hamiltonian, and the Gauss law and its imposition (Fig.~\ref{fig:model}). Later
sections use these conventions without redefinition.

\subsection{Lattice, sizes, and plaquettes}\label{sec:lattice}

We consider a two-dimensional square lattice with open boundaries, with $L_x$
columns and $L_y$ rows, sites $v=(x,y)$ with $x=0,\dots,L_x-1$ and
$y=0,\dots,L_y-1$, and links $l=((x,y),(x+1,y))$ (horizontal) or
$l=((x,y),(x,y+1))$ (vertical). Throughout this paper $L_x=2$ and $L_y\in\{2,3,4,5\}$, so the ladders are
$2\times2$--$2\times5$, with $L_y$ the row count and second
label index. Two $L_x=3$ pilots are discussed in Section~\ref{sec:discussion}
and Appendix~\ref{app:scaling}: $3\times2$ ($N_f=6$, even) and half-filled
$3\times3$ ($N_f=9$, odd). The numbers of links and plaquettes are
\begin{equation}
n_L=(L_x-1)L_y+L_x(L_y-1),\qquad n_p=(L_x-1)(L_y-1).
\end{equation}
A plaquette $p$ with lower-left corner $(x,y)$ consists of the four links
\begin{equation}
l_0:(x,y)\to(x+1,y),\quad
l_1:(x+1,y)\to(x+1,y+1),\quad
l_2:(x,y+1)\to(x+1,y+1),\quad
l_3:(x,y)\to(x,y+1),
\label{eq:plaqorder}
\end{equation}
taken counter-clockwise; the first two are traversed in their positive direction
and the last two opposite to it.

\subsection{The five-state link}\label{sec:link}

Each link carries the five-state $\{5\}$ truncation of the SU(2) quantum link
model \cite{Ludwig2026}. The exact link Hilbert space is
$\bigoplus_j V_j\otimes V_j^{*}$, where the left and right rishons sit at the two
ends of the link; the $\{5\}$ truncation keeps the $j=\tfrac12$ sector
($2\times2=4$ states) together with the $j=0$ singlet, so that
\[
\mathcal H_l=\underbrace{\mathrm{span}\{\ket{0,0}\}}_{j=0,\ 1\ \text{state}}
\;\oplus\;
\underbrace{\mathrm{span}\{\ket{\alpha,\beta}\}_{\alpha,\beta=0,1}}_{j=1/2,\ 4\ \text{states}},
\qquad \dim\mathcal H_l=5 .
\]
Here $\alpha,\beta\in\{0,1\}$ label the colour of the left and right rishon,
respectively; $\ket{0,0}$ is the rishon vacuum and $\ket{\alpha,\beta}$ has exactly one
rishon at each end. We write the orthonormal basis as $\ket{1},\dots,\ket{5}$ with
$\ket{5}\equiv\ket{0,0}$, and the four remaining states are the
$\ket{\alpha,\beta}$ in the order fixed below.

With $\ket{\alpha,\beta}=\ket{\alpha}_L\otimes
\ket{\beta}_R$,
\begin{equation}
E_L^a=\tfrac{\sigma^a}{2}\otimes\mathbf{1},\qquad
E_R^a=\mathbf{1}\otimes\tfrac{\sigma^a}{2}\qquad(\text{on the }j=\tfrac12\ \text{sector}),
\label{eq:elink}
\end{equation}
and $E_L^a\ket{5}=E_R^a\ket{5}=0$. These are the only definitions needed to
evaluate the algebra
\[
[E_L^a,E_L^b]=i\epsilon^{abc}E_L^c,\qquad
[E_R^a,E_R^b]=i\epsilon^{abc}E_R^c,\qquad
[E_L^a,E_R^b]=0,
\]
the Casimir
\begin{equation}
E^2=\sum_{a=1}^{3}(E_L^a)^2=\sum_{a=1}^{3}(E_R^a)^2
=\operatorname{diag}\!\left(\tfrac34,\tfrac34,\tfrac34,\tfrac34,0\right),
\label{eq:casimir}
\end{equation}
and the transformation properties of the link field,
\[
[E_L^a,U_{\alpha\beta}]=-(T^a)_{\alpha\gamma}U_{\gamma\beta},\qquad
[E_R^a,U_{\alpha\beta}]=U_{\alpha\gamma}(T^a)_{\gamma\beta},
\qquad T^a=\tfrac{\sigma^a}{2},
\]
which state that $U_{\alpha\beta}$ transforms in the fundamental representation at
each end. In the rishon representation $U$ creates or destroys a rishon pair, so
its only non-zero matrix elements connect the four $j=\tfrac12$ states with the
singlet:
\begin{align}
U_{00}&=-\sqrt2\,\ket{3}\!\bra{5}+\sqrt2\,\ket{5}\!\bra{2},&
U_{01}&=-\sqrt2\,\ket{4}\!\bra{5}-\sqrt2\,\ket{5}\!\bra{1},\\
U_{10}&=\sqrt2\,\ket{1}\!\bra{5}+\sqrt2\,\ket{5}\!\bra{4},&
U_{11}&=\sqrt2\,\ket{2}\!\bra{5}-\sqrt2\,\ket{5}\!\bra{3},
\label{eq:Umatrix}
\end{align}
with $U^\dagger_{\alpha\beta}=(U_{\alpha\beta})^\dagger$; every entry has modulus
$\sqrt2$, which is the normalization of the $\{5\}$ truncation. The resulting
basis labelling is summarized in Table~\ref{tab:link}.

\begin{table}[H]
\centering
\footnotesize
\caption{The five link states: rishon content, colour labels, and the
electric quantum numbers. The second column gives the rishon numbers
$(r_L,r_R)$ at the left and right ends of the link: $r_L=r_R=1$ for the four
$j=\tfrac12$ states and $r_L=r_R=0$ for the singlet. The third column gives the
colour labels $(\alpha,\beta)$ of the left and right rishons and is read off
\eqref{eq:Umatrix}, using
$U_{\alpha\beta}\ket{5}\propto\ket{1-\alpha,\beta}$ (the left index is
complemented, because the link field carries an antifundamental index at one
end); the fourth and fifth columns are the eigenvalues of $E_L^3$ and $E_R^3$
implied by \eqref{eq:elink}.}
\label{tab:link}
\begin{tabular}{cccrrrc}
\toprule
state & $(r_L,r_R)$ & $(\alpha,\beta)$ & $E_L^3$ & $E_R^3$ & $E^2$ & produced from $\ket{5}$ by \\
\midrule
$\ket{1}$ & $(1,1)$ & $(0,0)$ & $+\tfrac12$ & $+\tfrac12$ & $\tfrac34$ & $U_{10}$ \\
$\ket{2}$ & $(1,1)$ & $(0,1)$ & $+\tfrac12$ & $-\tfrac12$ & $\tfrac34$ & $U_{11}$ \\
$\ket{3}$ & $(1,1)$ & $(1,0)$ & $-\tfrac12$ & $+\tfrac12$ & $\tfrac34$ & $U_{00}$ \\
$\ket{4}$ & $(1,1)$ & $(1,1)$ & $-\tfrac12$ & $-\tfrac12$ & $\tfrac34$ & $U_{01}$ \\
$\ket{5}$ & $(0,0)$ & --- & $0$ & $0$ & $0$ & --- \\
\bottomrule
\end{tabular}
\end{table}

\subsection{Fermions}\label{sec:fermions}

Each lattice site hosts one staggered fermion flavour with two colour components,
so the local Hilbert space is four-dimensional. We use the ordered basis
\[
\ket{0},\quad \ket{1}\equiv c^\dagger_0\ket{0},\quad
\ket{2}\equiv c^\dagger_1\ket{0},\quad
\ket{12}\equiv c^\dagger_0c^\dagger_1\ket{0},
\]
i.e.\ $\ket{n_0,n_1}$ with the correspondence
$\ket{0}\leftrightarrow\ket{0,0}$,
$\ket{1}\leftrightarrow\ket{1,0}$,
$\ket{2}\leftrightarrow\ket{0,1}$, and
$\ket{12}\leftrightarrow\ket{1,1}$. The operators are
\begin{equation}
c_0=\tau^-\otimes\mathbf{1},\qquad
c_1=\sigma_z\otimes\tau^-,\qquad
\tau^-=\ket{0}_{\!f}\!\bra{1}_{\!f}=\begin{pmatrix}0&1\\0&0\end{pmatrix},
\label{eq:clocal}
\end{equation}
where $\ket{0}_{\!f},\ket{1}_{\!f}$ are the single-mode states (distinct from the
four site states above), with $\tau^+=\ket{1}_{\!f}\!\bra{0}_{\!f}=(\tau^-)^\dagger$
and $\ket{0}_{\!f}=(1,0)^T$, $\ket{1}_{\!f}=(0,1)^T$. The factor $\sigma_z$ in $c_1$ is the
Jordan--Wigner string of the first colour at the same site: the two colours must
anticommute, and $\sigma_z\tau^+=-\tau^+\sigma_z$ gives
$\{c_0,c_1\}=0$ while $\{c_0,c^\dagger_0\}=\{c_1,c^\dagger_1\}=1$. The colour
charge and fermion number of a site are
\[
Q_v^a=c^\dagger_{v,\alpha}T^a_{\alpha\beta}c_{v,\beta},\qquad
n_v=c^\dagger_{v,\alpha}c_{v,\alpha},
\]
where $\alpha,\beta\in\{0,1\}$ and $a=1,2,3$, so that $Q_v^a$ is the generator of
colour rotations on the site and $n_v$ counts
the fermions there.

\subsection{Hamiltonian}\label{sec:hamiltonian}

The Hamiltonian is $H=H_{\rm gauge}+H_{\rm hop}+H_{\rm mass}$, with
\begin{align}
H_{\rm gauge}
&=\frac{g^2}{2}\sum_{l}E_l^2
-\frac{1}{2g^2}\sum_{p}\left(P_p+P_p^\dagger\right),
\label{eq:hgauge}\\
H_{\rm hop}
&=-t\sum_{\langle v,w\rangle}\eta_{v,w}
\left(c^\dagger_{v,\alpha}(U_{v,w})_{\alpha\beta}c_{w,\beta}+\text{h.c.}\right),
\\
H_{\rm mass}
&=m\sum_v(-1)^{x_v+y_v}n_v .
\label{eq:hmass}
\end{align}
We write $H_{\rm gauge}=H_{\rm el}+H_{\rm mag}$ with
$H_{\rm el}=\frac{g^2}{2}\sum_l E_l^2$ and
$H_{\rm mag}=-\frac{1}{2g^2}\sum_p(P_p+P_p^\dagger)$.
Here $t$ is the hopping amplitude, $g$ is the gauge coupling, $m$ is the
staggered mass, and $(-1)^{x_v+y_v}$ is the staggered mass sign; all energies are
quoted in units of $t$. $E_l^2$ is the Casimir \eqref{eq:casimir} on link $l$,
and $U_{v,w}$ is the link operator of the link joining $v$ and $w$ with the
colour indices contracted as written. The plaquette operator is the colour trace
\begin{equation}
P_p=\sum_{i=0}^{1}\left(U_{l_0}U_{l_1}U^\dagger_{l_2}U^\dagger_{l_3}\right)_{ii},
\label{eq:plaquette}
\end{equation}
where the four links of the plaquette with lower-left corner $(x,y)$ are taken in
the order \eqref{eq:plaqorder}, i.e.\ counter-clockwise; the first two are
traversed in their positive direction and enter with $U$, while $l_2$ and $l_3$
are traversed opposite to their stated orientation and enter with $U^\dagger$.
Each $U_l$ in \eqref{eq:plaquette} is a $5\times5$ link operator carrying
$2\times2$ colour indices, so \eqref{eq:plaquette} is a sum of $2^{4}=16$ colour
monomials.

We use the standard two-dimensional Kogut--Susskind
staggering \cite{KogutSusskind1975}. For a bond $\langle v,w\rangle$ we take $v$
to be the left/lower endpoint and set
\begin{equation}
\eta_{v,w}=
\begin{cases}
1, & w=v+\hat x \quad(\text{horizontal}),\\[2pt]
(-1)^{x_v}, & w=v+\hat y \quad(\text{vertical}),
\end{cases}
\label{eq:stagger}
\end{equation}
i.e.\ $\eta_x(v)=1$ and $\eta_y(v)=(-1)^{x_v}$, the usual direction-dependent
choice.

\subsection{Gauss law and its imposition}\label{sec:gauss}

The Gauss generators at site $v$ are
\begin{equation}
G_v^a=\sum_{l\ni v}E_{l,v}^a+Q_v^a,
\qquad
E_{l,v}^a=
\begin{cases}
E_L^a,& v \text{ is the left/lower endpoint of } l,\\
E_R^a,& v \text{ is the right/upper endpoint of } l,
\end{cases}
\label{eq:gauss}
\end{equation}
and physical states satisfy
\begin{equation}
G_v^a\ket{\psi}=0\qquad\text{for all } v,a .
\label{eq:gausslaw}
\end{equation}
The relative sign between the $E_L$ and $E_R$ contributions in \eqref{eq:gauss} is
the one for which $[H,G_v^a]=0$.

Neither method enforces \eqref{eq:gausslaw} exactly. Both add the penalty
\begin{equation}
H_\lambda=H+\lambda\sum_{v,a}\left(G_v^a\right)^2,
\qquad \lambda=10 ,
\label{eq:penalty}
\end{equation}

\section{Methods}\label{sec:methods}

Two independent Hilbert-space methods are used. Sparse ED, available only on the
$2\times2$ lattice, is the reference; DMRG is the scalable method and covers
all five geometries. The $2\times2$ comparison tests the DMRG machinery,
especially the Jordan--Wigner convention, before the larger ladders are trusted.

Both methods are implemented in Python: the sparse ED uses NumPy and SciPy's
Lanczos solver \cite{SciPy}; the DMRG uses TeNPy 1.1.1 \cite{TeNPy}, with the
MPO built term by term on top of TeNPy's \texttt{CouplingModel}. The independent
fPEPS cross-check of Section~\ref{sec:discussion} uses quimb and symmray. Each
colour mode is a \texttt{FermionSite(conserve="N")} and each link a five-state
site with trivial charge, so the Jordan--Wigner string is supplied by TeNPy's
fermionic site. The component comparison of Section~\ref{sec:validation}
validates the custom MPO construction, not the TeNPy algorithms themselves.

\subsection{Sparse exact diagonalization}\label{sec:ed}

On the $2\times2$ lattice the product basis has dimension
$5^{4}\cdot4^{4}=160{,}000$. All horizontal links are enumerated first in
row-major order, followed by all vertical links in row-major order, and this
ordering defines the product basis. A basis state is the tuple
\[
(\text{link}_0,\dots,\text{link}_{n_L-1},\ \text{site}_0,\dots,\text{site}_{n_V-1}),
\]
with each link index running over the five states of Table~\ref{tab:link} and each
site index over the four states of Section~\ref{sec:fermions}; $n_V=L_xL_y$ is the
number of sites. The Hamiltonian is assembled as a complex sparse (CSR) matrix from
Kronecker products of local operators, one term of \eqref{eq:hgauge}--\eqref{eq:hmass}
at a time, and the penalty \eqref{eq:penalty} is assembled in the same way.

\paragraph{Jordan--Wigner convention.} Modes are ordered site by site and, within
a site, colour by colour, consistent with the local basis \eqref{eq:clocal}. Let
$\sigma(v)$ denote the row-major position of site $v$ ($y$ outer, $x$ inner).
The ED uses this site ordering; the MPS chain of Section~\ref{sec:mps}
interleaves link sites with it, but link sites carry trivial parity and do not
change the string. On this ordering the Jordan--Wigner transformation is
\[
c_j=\Bigl(\prod_{k<j}Z_k\Bigr)\sigma^-_j,\qquad
c^\dagger_j=\Bigl(\prod_{k<j}Z_k\Bigr)\sigma^+_j,
\]
where $j$ runs over the $2L_xL_y$ colour modes and
$Z_k=(-1)^{n_k}=1-2n_k=\operatorname{diag}(1,-1)$ is the parity matrix in the
empty/occupied basis. The strings of $c^\dagger$ and $c$ cancel outside the
interval between the two modes, so a hopping term carries the product of $Z_k$
over the intermediate modes. Grouping the two colour modes of a site gives the
site parity
\[
\Pi_u=Z_{u,0}Z_{u,1}=(-1)^{n_{u,0}+n_{u,1}}=(-1)^{n_u},
\qquad n_{u,\alpha}=c^\dagger_{u,\alpha}c_{u,\alpha},
\]
which is $\operatorname{diag}(1,-1,-1,1)$ in the ordered basis
$\ket0,\ket1,\ket2,\ket{12}$, i.e.\ $+1$ for $n_u=0$ or $2$ and $-1$ for
$n_u=1$; the on-site $\sigma_z$ in \eqref{eq:clocal} accounts for the order of
the two colours. For a hopping term on a bond $\langle v,w\rangle$ the operator
inserted in the ED matrix is
\begin{equation}
O_{v\to w}=c^\dagger_{v,\alpha}
\Bigl(\prod_{\sigma(v)<\sigma(u)<\sigma(w)}\Pi_u\Bigr)
(U_{v,w})_{\alpha\beta}\,c_{w,\beta}.
\label{eq:jw}
\end{equation}
The product runs over all sites strictly between $v$ and $w$ in this
one-dimensional ordering, not over the spatial path, and the string is attached
to the $v$ side, i.e.\ to the creation operator. The Hermitian conjugate is
obtained by exchanging $v\leftrightarrow w$ and replacing $U$ by $U^\dagger$;
the string is Hermitian. Here $(U_{v,w})_{\alpha\beta}$ is the single link
operator on the bond $\langle v,w\rangle$; it is not a Wilson line. Products of
link operators appear only in the plaquette term $P_p$. The same convention is
used in the MPO (Section~\ref{sec:mpo}); the component comparison of
Section~\ref{sec:validation} is the check of \eqref{eq:jw}.

Hopping terms are assembled link by link: for each colour pair $(\alpha,\beta)$
we insert $c^\dagger_{v,\alpha}$ on the left site $v$, $U_{\alpha\beta}$ on the
bond, and $c_{w,\beta}$ on the right site $w$, with $\Pi_u$ on every
intermediate site; the left-end factor is ordered as $c^\dagger_{v,\alpha}\Pi_v$
in the forward term and $\Pi_v c_{v,\alpha}$ in the Hermitian conjugate, so the
pair is Hermitian. The magnetic term is the sum of the $2^4$ colour monomials of
\eqref{eq:plaquette}.

The ground state of $H+H_\lambda$ is obtained by sparse diagonalization, without
any constraint on the fermion number or on \eqref{eq:gausslaw}. The returned
state is therefore a physical state only up to the penalty, which is why the
Gauss residual is measured on the eigenvector rather than assumed.

\subsection{Lattice to MPS chain}\label{sec:mps}

For DMRG the two-dimensional lattice is mapped to a one-dimensional MPS chain of
\textbf{heterogeneous} sites: fermion modes and link degrees of freedom are
interleaved. For each row $y$ we place, from left to right, the two colour modes of
each site followed by the horizontal link to the right of that site, and after the
row we append the vertical links of that row in increasing $x$, i.e.\
\begin{equation}
\underbrace{\bigl[s_{0,y}^{(c_0)},s_{0,y}^{(c_1)},h_{0,y}\bigr]\bigl[s_{1,y}^{(c_0)},s_{1,y}^{(c_1)},h_{1,y}\bigr]\cdots\bigl[s_{L_x-1,y}^{(c_0)},s_{L_x-1,y}^{(c_1)}\bigr]}_{\text{row } y,\ \text{horizontal links}},\;
\underbrace{v_{0,y},\dots,v_{L_x-1,y}}_{\text{vertical links of row } y\ (y<L_y-1)},
\label{eq:chain}
\end{equation}
repeated for $y=0,\dots,L_y-1$. Here $h_{x,y}$ is the horizontal link
$(x,y)\to(x+1,y)$, present for $x=0,\dots,L_x-2$; $v_{x,y}$ is the vertical link
$(x,y)\to(x,y+1)$, present for $x=0,\dots,L_x-1$ and $y=0,\dots,L_y-2$. The number
of MPS sites is therefore
\[
N_{\rm MPS}=2L_xL_y+n_L.
\]
This ordering is not a snake: every row is traversed
left to right. It keeps the four links of a plaquette within a window of
$4L_x$ consecutive MPS sites ($8$ sites for $L_x=2$), which is why the MPO
remains local; geometries with a wider cross-section ($L_x>2$) would need a
snake or cylinder ordering, as discussed in
Section~\ref{sec:discussion}.

Fermion modes are represented by sites that conserve the total fermion number, so
that $N$ is an exact quantum number of the MPS. Link sites have local dimension
five and carry a trivial charge. Because $N$ is exact, the fermion number of a DMRG run is fixed by the initial
state rather than by the solver. At half filling the expected value is
$N_f=L_xL_y$.

\subsection{MPO construction}\label{sec:mpo}

The MPO is built from local terms. The Jordan--Wigner strings required by the
fermion operators are inserted so that the MPO convention matches \eqref{eq:jw}
by construction, and products of operators acting on the same site are handled
explicitly. Each term of the Hamiltonian is added as a sum of local monomials.

\begin{itemize}
\item \emph{Electric term.} $\frac{g^2}{2}E_l^2$ is a one-site operator on each
link site, diagonal in the basis of Table~\ref{tab:link}.
\item \emph{Magnetic term.} The trace \eqref{eq:plaquette} is added as the $2^4$
colour monomials $U_{ij}U_{jk}U^\dagger_{kl}U^\dagger_{li}$, $i,j,k,l\in\{0,1\}$,
each a product of four link operators, together with the Hermitian conjugate and
coefficient $-1/(2g^2)$. Each monomial spans the four links of a plaquette.
\item \emph{Hopping term.} $-t\eta_{v,w}c^\dagger_{v,\alpha}
(U_{v,w})_{\alpha\beta}c_{w,\beta}$ plus its Hermitian conjugate, with
$\eta_{v,w}$ from \eqref{eq:stagger}; the colour indices are contracted before the
term is added, so each hopping term is a sum of four monomials.
\item \emph{Mass term.} Diagonal in the fermion number, one site at a time.
\item \emph{Gauss penalty.} Expanded into monomials before being added, see below.
\end{itemize}

\paragraph{Expansion of the Gauss penalty.} Squaring \eqref{eq:gauss} gives four
structurally different classes of terms on a site $v$:
\[
\Bigl(\sum_{l\ni v}E_{l,v}^a+Q_v^a\Bigr)^{2}
=\underbrace{\sum_{l,l'\ni v}E^a_{l,v}E^a_{l',v}}_{\text{link--link (2 sites at most)}}
\;+\;2\underbrace{\sum_{l\ni v}E^a_{l,v}Q_v^a}_{\text{link--fermion (2 sites)}}
\;+\;\underbrace{Q_v^aQ_v^a}_{\text{fermion only (1 site)}} .
\]
The first class contains on-site ($l=l'$) and next-neighbour ($l\neq l'$) pieces,
the second couples a link site to the fermion site at the same vertex, and the
third is a single-site two-fermion operator; summing over $a$ and using the SU(2)
completeness relation gives
\[
\sum_{a=1}^{3}Q_v^aQ_v^a=\frac32\,n_v-\frac34\,n_v^2 ,
\]
which is used to reduce the colour part of the penalty to $n_v$ and $n_v^2$. The
sum over $a$ is taken before the monomials are added. This expansion is the step
where an implementation error is easiest to make, because the link--fermion class
mixes two different site types and requires the correct site order.

The resulting MPO bond dimension is reported in Table~\ref{tab:mpo}; it reaches
about $40$ on the $2\times4$ and $2\times5$ ladders and $71$ on the $3\times3$
pilot, and the bond dimension of the MPS, rather than that of the MPO, is what
limits the ladder length. The MPO is Hermitian by construction, since every term is added together with its
adjoint.

\begin{table}[H]
\centering
\footnotesize
\caption{MPO size versus MPS size. The MPO bond dimension is independent of the
MPS bond dimension and grows only mildly, from 32 on $2\times2$ to about 40 on
the ladders and 71 on $3\times3$; for the ladders the limiting resource is the
MPS bond dimension $\chi$, not the MPO.}
\label{tab:mpo}
\begin{tabular}{lcc}
\toprule
geometry & $N_{\rm MPS}$ & max MPO bond dimension \\
\midrule
$2\times2$ & 12 & 32 \\
$2\times3$ & 19 & 40 \\
$2\times4$ & 26 & 40 \\
$2\times5$ & 33 & 40 \\
$3\times3$ & 30 & 71 \\
\bottomrule
\end{tabular}
\end{table}

\subsection{DMRG}\label{sec:dmrg}

The ground state of $H_\lambda$ is computed with the two-site DMRG algorithm
\cite{Schollwock2011} on the chain \eqref{eq:chain}. Two-site sweeps with a finite
bond dimension converge to a local minimum of the variational functional
restricted to the rank-$\chi$ manifold, which is not in general the rank-$\chi$
projection of the ground state. This is why the acceptance rules below
require more than one initial state. For every run we
record the number of sweeps, the final sweep-to-sweep energy change, and the
maximum truncation error; the accepted points are those whose energy is stable
under a further increase of $\chi$.

\paragraph{Initial states.} All link sites start in the rishon vacuum $\ket{5}$.
The fermion product states are specified by which of the $2L_xL_y$ colour modes
are occupied; each state contains exactly $N_f=L_xL_y$ fermions, i.e.\ it is half
filled.
\begin{itemize}
\item \texttt{color1}: all colour-0 modes occupied, all colour-1 modes empty;
\item \texttt{color2}: all colour-1 modes occupied, all colour-0 modes empty;
\item \texttt{mixed}: on each site $(x,y)$ the colour $c\in\{0,1\}$ with
  $x+y+c$ even is occupied and the other colour is empty;
\item \texttt{double}: both colours are occupied on the first $L_xL_y/2$ sites
  and both are empty on the remaining sites (defined for even $n_V$; not used
  for the $3\times3$ pilot);
\item \texttt{random}: exactly $N_f$ of the $2L_xL_y$ colour modes are chosen
  uniformly at random.
\end{itemize}
These states span different filling patterns and are used to detect local minima;
the comparison is not intended to accelerate convergence.

\paragraph{Observables.} After each run we measure the energy $E_0$, the total
fermion number $N_f=\sum_v\langle n_v\rangle$, the Gauss residual
$\langle G^2\rangle=\sum_{v,a}\langle(G_v^a)^2\rangle$, and the normalized
plaquette $\langle P_p\rangle/16$, normalized so that the pure-gauge plaquette
approaches unity as $g\to0$. The latter is extracted from the magnetic part of
the MPO evaluated on the final MPS,
\[
\frac{\langle P_p\rangle}{16}=-\frac{g^2}{16\,n_p}\langle H_{\rm mag}\rangle,
\]
which follows from \eqref{eq:hgauge}.

\subsection{Convergence and acceptance rules}\label{sec:control}

The ED reference is exact up to numerical precision and needs no acceptance rule. For DMRG a point is accepted only if all four conditions hold.

\begin{enumerate}
\item \emph{Bond-dimension stability}: $|E(\chi)-E(\chi')|\le\epsilon_E$ for the
two largest available bond dimensions $\chi'<\chi$ at that geometry, with
$\epsilon_E=10^{-6}$.
\item \emph{Truncation error}: the largest truncation error over the final sweep is
below $\epsilon_{\rm trunc}=10^{-6}$.
\item \emph{Gauss law}: $\langle G^2\rangle\le10^{-8}$.
\item \emph{Landscape check}: the lowest energy found across all initial states is
not lowered by any other initial state. If another initial state finds a lower
energy, the point is reclassified and the lower value is used.
\end{enumerate}

Conditions 1--3 alone are not sufficient, and this is the methodological point of
the paper: condition 1 is evaluated \emph{within one run} and therefore cannot
detect a local minimum, condition 2 measures the truncation of that run only, and
condition 3 can be satisfied by an unphysical state if the penalty is large enough.
Condition 4 is the only one that tests the landscape rather than the run, and it is
the rule that separates a converged point from a trapped one. Points failing any
condition are reported explicitly as not converged and are excluded from the
reliability envelope; they are not silently dropped.

The energy error has three contributions: bond-dimension truncation, estimated by
$|E(\chi)-E(\chi')|$ (condition 1); the finite penalty, checked by repeating the
$2\times3$ point at $\lambda=100$ (Appendix~\ref{app:params}); and the
initial-state landscape (condition 4). The value $\lambda=10$ suppresses
$\langle G^2\rangle$ below $10^{-8}$ for the accepted points while keeping
$H_\lambda$ well conditioned. For $2\times2$ the first contribution is at the
$10^{-13}$ level (ED cross-check); for $2\times3$ the truncation contributes at
$10^{-6}$ and the penalty below $10^{-5}$; for $2\times4$ and $2\times5$ the
truncation contributes at $10^{-5}$, with the Gauss residual below
$2.3\times10^{-9}$. For $3\times2$ none of the three is controlled, which is why
it is excluded from the envelope.

\section{Validation}\label{sec:validation}

\subsection{Exact identities}

The link algebra is checked directly against the defining commutators. For the
full Hamiltonian we verify $[H,G_v^a]=0$ and $H=H^\dagger$ numerically. At
$t=m=0$ the model reduces to pure gauge theory; at $g=1$ the $2\times2$ ground
energy is $-15.26756848$, which we use as a regression test.

\subsection{Component comparison}

At $g=1$, $m=0.1$ we compare the MPO expectation values of the electric,
magnetic, mass, hopping, gauge, and penalty terms with the sparse ED. The
hopping term is the most sensitive to the JW convention. After correcting the
cross-site JW strings, all components agree at the $10^{-14}$ level; the full
energy is
\[
E_0^{\rm MPO}=-19.413260195978,\qquad
E_0^{\rm ED}=-19.413260195978.
\]
This is a strong check that the two implementations represent the same
fermionic model, not merely the same bosonic truncation. Table~\ref{tab:components}
lists the full component-by-component comparison; the penalty MPO is checked
separately on product states in Appendix~\ref{app:params}.

\begin{table}[H]
\centering
\footnotesize
\caption{Component-by-component comparison of the MPO and ED at
$g=1$, $m=0.1$ on $2\times2$ (energies in units of $t$). The last column is the
absolute difference.}
\label{tab:components}
\begin{tabular}{lccc}
\toprule
component & MPO & ED & $|{\rm diff}|$ \\
\midrule
electric & $0.656517131993$ & $0.656517131993$ & $2.2\times10^{-16}$ \\
magnetic & $-12.178640504104$ & $-12.178640504104$ & $1.8\times10^{-15}$ \\
mass & $-0.066617657029$ & $-0.066617657029$ & $1.1\times10^{-15}$ \\
hopping & $-7.824519166839$ & $-7.824519166839$ & $8.9\times10^{-15}$ \\
gauge & $-11.522123372110$ & $-11.522123372110$ & $5.3\times10^{-15}$ \\
penalty & $\sim0$ & $\sim0$ & $2.3\times10^{-15}$ \\
full & $-19.413260195978$ & $-19.413260195978$ & $0$ \\
\bottomrule
\end{tabular}
\end{table}

\subsection{\texorpdfstring{$2\times2$}{2x2} ED versus DMRG}

Table~\ref{tab:2x2} lists the fixed points used in this paper. The DMRG and ED
energies agree to $10^{-13}$ or better, and the Gauss residual is below
$10^{-15}$. These points form the exact reference set for the method
benchmark.

\begin{table}[H]
\centering
\footnotesize
\caption{$2\times2$ reliability set ($t=1$, $\lambda=10$). The DMRG value is the
lowest energy found across the initial states of Section~\ref{sec:dmrg}.}
\label{tab:2x2}
\begin{tabular}{ccrrrrr}
\toprule
$g$ & $m$ & $E_0$ (DMRG) & $E_0$ (ED) & $|E_{\rm DMRG}-E_{\rm ED}|$ &
plaquette & $G^2$ \\
\midrule
0.5 & 0.1 & -64.926475062 & -64.926475062 & $8.5\times10^{-14}$ & 0.987283 & $<10^{-15}$ \\
1.0 & 0.1 & -19.413260196 & -19.413260196 & $3.2\times10^{-14}$ & 0.761165 & $<10^{-15}$ \\
2.0 & 0.1 & -10.220071849 & -10.220071849 & $4.3\times10^{-14}$ & 0.452055 & $<10^{-15}$ \\
1.0 & 0.0 & -19.379644646 & -19.379644646 & $7.5\times10^{-14}$ & 0.757633 & $<10^{-15}$ \\
1.0 & 0.2 & -19.510627935 & -19.510627935 & $3.9\times10^{-14}$ & 0.770680 & $<10^{-15}$ \\
\bottomrule
\end{tabular}
\end{table}

\section{Results}\label{sec:results}

\subsection{The \texorpdfstring{$2\times3$}{2x3} lattice}

\paragraph{Coupling dependence.} Table~\ref{tab:2x3g} gives the converged $2\times3$ results at $m=0.1$. The
Gauss residual is below $10^{-11}$ in all cases. The plaquette decreases with
increasing $g$, while the energy scale changes strongly because the magnetic
term scales as $1/g^2$.

\begin{table}[H]
\centering
\footnotesize
\caption{$2\times3$ results at $m=0.1$ ($t=1$, $\lambda=10$). The quoted value
at each $g$ is the converged energy; all sampled initial states agree.}
\label{tab:2x3g}
\begin{tabular}{crrr}
\toprule
$g$ & $E_0$ & plaquette & $G^2$ \\
\midrule
0.5 & -103.579704 & 0.773325 & $2.6\times10^{-12}$ \\
1.0 & -32.699014 & 0.627592 & $3.4\times10^{-13}$ \\
2.0 & -16.490333 & 0.437372 & $3.7\times10^{-14}$ \\
\bottomrule
\end{tabular}
\end{table}

For the $2\times3$ points the energy is stable between
the two largest bond dimensions at the $10^{-6}$ level, the truncation errors
are at most $10^{-7}$, and the Gauss residuals are at most $10^{-11}$. The
dominant residual uncertainty is therefore the bond-dimension extrapolation,
not the penalty or the truncation; the $\lambda=100$ repeat of
Appendix~\ref{app:params} confirms this.

\paragraph{Mass dependence.} The mass scan at $g=1$ (Table~\ref{tab:2x3m}) is under control: $m=0$ gives
$-32.655703$ at both $\chi=384$ and $\chi=448$ (agreeing to $10^{-10}$); $m=0.2$
gives $-32.824271$ with
$G^2=1.3\times10^{-13}$. The $m=0.1$ point is the one discussed above.

\begin{table}[H]
\centering
\footnotesize
\caption{$2\times3$ mass scan at $g=1$ ($t=1$, $\lambda=10$). The $m=0.1$ row is
the same point as in Table~\ref{tab:2x3g}.}
\label{tab:2x3m}
\begin{tabular}{crrr}
\toprule
$m$ & $E_0$ & plaquette & $G^2$ \\
\midrule
0.0 & -32.655703 & 0.625842 & $1.7\times10^{-13}$ \\
0.1 & -32.699014 & 0.627592 & $3.4\times10^{-13}$ \\
0.2 & -32.824271 & 0.632301 & $1.3\times10^{-13}$ \\
\bottomrule
\end{tabular}
\end{table}

\paragraph{Initial-state insensitivity.} The initial-state dependence is weak. At $g=1$, $m=0.1$, the \texttt{color1},
\texttt{mixed}, and \texttt{random} states all converge to $E_0=-32.699014$;
the same holds at $g=0.5$ ($-103.579704$) and $g=2$ ($-16.490333$). The
landscape check of Section~\ref{sec:control} remains part of the acceptance
protocol, and it is satisfied by the accepted $2\times3$ points. The
corresponding mixed/random agreement on the $2\times4$ and $2\times5$ ladders
is reported in Section~\ref{sec:large-ladders}.

\subsection{The \texorpdfstring{$2\times4$ and $2\times5$}{2x4 and 2x5} ladders}\label{sec:large-ladders}

\begin{figure}[H]
\centering
\includegraphics[width=0.62\linewidth]{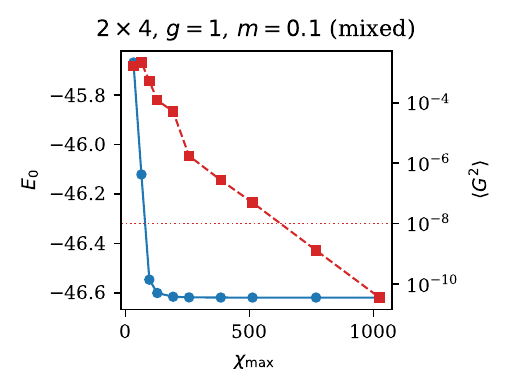}
\caption{$2\times4$ convergence at $g=1$, $m=0.1$ with the \texttt{mixed}
initial state. The energy (left axis, circles) is stable at the $10^{-5}$ level
between $\chi=768$ and $1024$; the Gauss residual (right axis, squares,
logarithmic) reaches $3.5\times10^{-11}$ at $\chi=1024$, below the acceptance
threshold $\langle G^2\rangle=10^{-8}$ (dotted line).}
\label{fig:2x4}
\end{figure}

The $2\times4$ MPO has 26 MPS sites and maximum MPO bond dimension about 40;
the $2\times5$ ladder has 33 MPS sites. At $\chi=1024$ the
two ladders give the results in Table~\ref{tab:2x45}. The energies change by
$2.8\times10^{-5}$ ($2\times4$) and $2.2\times10^{-5}$ ($2\times5$) between
$\chi=768$ and $1024$, so we quote them at the $10^{-5}$ level; the Gauss
residuals are $3.5\times10^{-11}$ and $2.3\times10^{-9}$, and the truncation
errors are $6.7\times10^{-10}$ and $1.7\times10^{-8}$. The mixed and random
initial states agree to $3\times10^{-8}$ ($2\times4$) and to $10^{-7}$
($2\times5$). The
dominant residual uncertainty for these ladders is the bond-dimension
extrapolation; the energy is stable at the $10^{-5}$ level, and the truncation
errors and Gauss residuals are already below the acceptance thresholds. The
convergence of the $2\times4$ energy and Gauss residual with $\chi$ is shown in
Fig.~\ref{fig:2x4}.

\begin{table}[H]
\centering
\footnotesize
\caption{$2\times4$ and $2\times5$ results at $g=1$, $m=0.1$ ($t=1$,
$\lambda=10$, $\chi=1024$). The quoted energy is stable at the $10^{-5}$ level
between $\chi=768$ and $1024$.}
\label{tab:2x45}
\begin{tabular}{lcccc}
\toprule
geometry & $E_0$ & plaquette & $G^2$ & $\epsilon_{\rm trunc}$ \\
\midrule
$2\times4$ & $-46.6192026$ & 0.616455 & $3.5\times10^{-11}$ & $6.7\times10^{-10}$ \\
$2\times5$ & $-60.3860082$ & 0.600949 & $2.3\times10^{-9}$ & $1.7\times10^{-8}$ \\
\bottomrule
\end{tabular}
\end{table}

\subsection{The \texorpdfstring{$3\times3$ and $3\times2$}{3x3 and 3x2} geometries}\label{sec:three}

Two $L_x=3$ pilots complete the envelope. The half-filled $3\times3$ lattice has
$N_f=9$, which is odd, so the total colour charge is half-integer; the links
carry integer colour charge, and the summed Gauss law cannot vanish. The
observed $G^2=3/4$ is the exact lower bound, independent of $\chi$ and
$\lambda$. With one hole or one doublon ($N_f=8$ or $10$) the residual falls to
$6\times10^{-4}$ at $\chi=512$, $1$--$2\times10^{-4}$ at $\chi=1024$, and
$1.6$--$2.5\times10^{-5}$ at $\chi=2048$; the truncation error is still
decreasing, so this residual is limited by $\chi$ rather than by the $3/4$
floor. The obstruction is therefore the filling parity, not the ordering or the
penalty.

The $3\times2$ lattice has $N_f=6$ (even) and admits an exact-Gauss physical
state, so $L_x=3$ is not by itself an obstruction. Its DMRG energy at $\chi=512$
($-32.5928$, $G^2=6.7\times10^{-11}$) lies above an independent exact-Gauss fPEPS
reference ($-32.61$ to $-32.64$)\footnote{The fPEPS uses explicit SU(2) singlet
projectors with virtual multiplicity $n=3$ and is a separate implementation; the
quoted range is the spread between initializations. It is not extrapolated in
$n$ and serves only as an independent lower reference; its construction will be
reported elsewhere.}; at $\chi=768$ the penalty-limited state ($E=-32.68$,
$G^2=1.1\times10^{-4}$) is unphysical. The $3\times2$ point therefore demonstrates
the even-$N_f$ sector but is not a converged energy benchmark.

Table~\ref{tab:envelope} collects the resulting reliability envelope.

\begin{table}[H]
\centering
\footnotesize
\caption{Reliability envelope. ED is used only on $2\times2$; all other points
are DMRG. The quoted error is the energy scale (in units of $t$) controlled by
the bond-dimension and penalty checks of Section~\ref{sec:control};
$\langle G^2\rangle$ is the Gauss residual.}
\label{tab:envelope}
\begin{tabular}{lccccl}
\toprule
geometry & $N_f$ & method & MPS $\chi$ & $\langle G^2\rangle$ & status \\
\midrule
$2\times2$ & 4 & ED+DMRG & 256 & $<10^{-15}$ & controlled ($10^{-13}$) \\
$2\times3$ & 6 & DMRG & 384--512 & $\le10^{-11}$ & controlled ($10^{-6}$) \\
$2\times4$ & 8 & DMRG & 1024 & $3.5\times10^{-11}$ & controlled ($10^{-5}$) \\
$2\times5$ & 10 & DMRG & 1024 & $2.3\times10^{-9}$ & controlled ($10^{-5}$) \\
$3\times2$ & 6 & DMRG & 512/768 & $6.7\times10^{-11}$ / $1.1\times10^{-4}$ & even filling; not converged \\
$3\times3$ & 9 & DMRG & 512--2048 & $3/4$ & obstructed (odd $N_f$) \\
$3\times3$ & 8/10 & DMRG & 2048 & $1.6$--$2.5\times10^{-5}$ & $\chi$-limited \\
\bottomrule
\end{tabular}
\end{table}

\section{Discussion and conclusions}\label{sec:discussion}

The benchmark establishes a reliable small-lattice envelope: $2\times2$ is
controlled by ED and DMRG to $10^{-13}$; the accepted $2\times3$ points are
controlled to the $10^{-6}$ level in energy and to $10^{-11}$ or better in
Gauss residual; and the $2\times4$ and $2\times5$ ladders are controlled to the
$10^{-5}$ level with Gauss residuals below $2.3\times10^{-9}$.
The dominant systematics are not statistical but structural: (i) the JW
convention must be identical in the ED and MPO implementations; (ii) the Gauss
penalty must be monitored separately, since a small energy can hide a
non-physical state; (iii) the initial state must still be varied to rule out local
minima, although the accepted $2\times3$ points and the $2\times4$/$2\times5$
ladders are insensitive to it.

The $3\times3$ and $3\times2$ pilots of Section~\ref{sec:three} complete the
envelope and expose the two failure modes that a residual alone cannot
distinguish: an exact symmetry obstruction (half-filled $3\times3$,
$G^2=3/4$) and a penalty-limited unconverged point ($3\times2$ at $\chi=768$).
A method benchmark must therefore fix and report the filling sector and the
penalty before attributing a residual to the ansatz. Together with the
cross-validated ED/DMRG envelope, Table~\ref{tab:envelope} provides a concrete
baseline for larger tensor-network or quantum-simulation studies.

The main open directions are exact Gauss projection at even $N_f$ or
gauge-invariant MPS initial states, a different MPS geometry such as a snake or cylinder ordering,
and an extension to larger ladders. Beyond the $L_x=2$ ladders, a
two-dimensional tensor network such as projected entangled pair states (PEPS), or an exact gauge-invariant
construction, is the natural route to larger $L_x$ \cite{Cataldi2024}. These are
required before the present approach can address continuum or phase-diagram
questions.

A three-dimensional extension, for example an open $2\times2\times L$ prism
mapped to a snake or cylinder chain, is a natural next step; its
two-dimensional limit ($L_y=1$) would provide a direct check against the
present code before any production run.

\section*{Acknowledgments}
We thank the Shanghai Jiao Tong University High-Performance Computing Center
(Siyuan-1, siyuan 1) for computational resources. D.~L.\ was supported by the
National Key Research and Development Program of China (No.~2024YFA1014104).

\appendix
\section{Numerical parameters and per-point diagnostics}\label{app:params}

Table~\ref{tab:diagnostics} lists the numerical diagnostics for the accepted
points and for the $3\times3$ pilot. For the $2\times2$ runs the DMRG value is
already controlled by the ED
reference, so only $G^2$ is quoted; for $2\times3$, $2\times4$, $2\times5$, and
$3\times3$ we give the bond dimension, the final sweep-to-sweep energy change,
the maximum truncation error, and the Gauss residual.

\begin{table}[H]
\centering
\footnotesize
\caption{Per-point DMRG diagnostics. $\Delta E_{\rm last}$ is the final
sweep-to-sweep energy change and $\epsilon_{\rm trunc}$ the maximum truncation
error over the final sweep. For the $2\times2$ points, which are controlled by
ED, these two columns are not needed (---). For the $3\times3$ rows,
\texttt{mixed} has $N_f=9$, while \texttt{hole} and \texttt{doublon} have
$N_f=8$ and $10$; the $3\times2$ rows have $N_f=6$.}
\label{tab:diagnostics}
\begin{tabular}{lccccccc}
\toprule
geometry & $g$ & $m$ & initial & $\chi$ & $\Delta E_{\rm last}$ &
$\epsilon_{\rm trunc}$ & $G^2$ \\
\midrule
$2\times2$ & 0.5 & 0.1 & color1 & 256 & --- & --- & $<10^{-15}$ \\
$2\times2$ & 1.0 & 0.1 & color1 & 256 & --- & --- & $<10^{-15}$ \\
$2\times2$ & 2.0 & 0.1 & color1 & 256 & --- & --- & $<10^{-15}$ \\
$2\times2$ & 1.0 & 0.0 & mixed  & 256 & --- & --- & $<10^{-15}$ \\
$2\times2$ & 1.0 & 0.2 & mixed  & 256 & --- & --- & $<10^{-15}$ \\
$2\times3$ & 0.5 & 0.1 & mixed  & 384 & $-5.4\times10^{-9}$ & $6.8\times10^{-9}$ & $2.6\times10^{-12}$ \\
$2\times3$ & 1.0 & 0.1 & random & 512 & $-6.9\times10^{-10}$ & $4.4\times10^{-20}$ & $3.4\times10^{-13}$ \\
$2\times3$ & 2.0 & 0.1 & mixed  & 384 & $-4.9\times10^{-10}$ & $2.7\times10^{-21}$ & $3.7\times10^{-14}$ \\
$2\times3$ & 1.0 & 0.2 & mixed  & 448 & $-2.5\times10^{-9}$ & $2.2\times10^{-21}$ & $1.3\times10^{-13}$ \\
$2\times3$ & 1.0 & 0.0 & mixed  & 448 & $-3.0\times10^{-9}$ & $2.5\times10^{-20}$ & $1.7\times10^{-13}$ \\
$2\times4$ & 1.0 & 0.1 & mixed  & 1024 & $-4.8\times10^{-9}$ & $6.7\times10^{-10}$ & $3.5\times10^{-11}$ \\
$2\times4$ & 1.0 & 0.1 & random & 1024 & $-5.5\times10^{-9}$ & $6.6\times10^{-10}$ & $3.6\times10^{-11}$ \\
$2\times5$ & 1.0 & 0.1 & mixed  & 1024 & $-1.4\times10^{-7}$ & $1.7\times10^{-8}$ & $2.3\times10^{-9}$ \\
$2\times5$ & 1.0 & 0.1 & random & 1024 & $-1.3\times10^{-7}$ & $1.8\times10^{-8}$ & $2.3\times10^{-9}$ \\
$3\times2$ & 1.0 & 0.1 & mixed  & 512 & $-1.3\times10^{-3}$ & $1.7\times10^{-7}$ & $1.9\times10^{-8}$ \\
$3\times2$ & 1.0 & 0.1 & random & 512 & $-9.5\times10^{-5}$ & $1.7\times10^{-7}$ & $6.7\times10^{-11}$ \\
$3\times2$ & 1.0 & 0.1 & mixed  & 768 & $-1.7\times10^{-2}$ & $1.5\times10^{-12}$ & $3.8\times10^{-7}$ \\
$3\times2$ & 1.0 & 0.1 & random & 768 & $-6.6\times10^{-3}$ & $9.4\times10^{-14}$ & $1.1\times10^{-4}$ \\
$3\times3$ & 1.0 & 0.1 & mixed  & 512 & $-2.6\times10^{-3}$ & $2.8\times10^{-4}$ & $0.75$ \\
$3\times3$ & 1.0 & 0.1 & hole    & 512  & $-5.9\times10^{-3}$ & $3.2\times10^{-4}$ & $6.2\times10^{-4}$ \\
$3\times3$ & 1.0 & 0.1 & hole    & 1024 & $-1.6\times10^{-4}$ & $1.3\times10^{-4}$ & $1.1\times10^{-4}$ \\
$3\times3$ & 1.0 & 0.1 & doublon & 512  & $-3.1\times10^{-4}$ & $5.2\times10^{-4}$ & $6.3\times10^{-4}$ \\
$3\times3$ & 1.0 & 0.1 & doublon & 1024 & $-1.2\times10^{-4}$ & $1.7\times10^{-4}$ & $2.2\times10^{-4}$ \\
$3\times3$ & 1.0 & 0.1 & hole    & 2048 & $-1.8\times10^{-4}$ & $1.4\times10^{-5}$ & $1.6\times10^{-5}$ \\
$3\times3$ & 1.0 & 0.1 & doublon & 2048 & $-8.9\times10^{-5}$ & $2.3\times10^{-5}$ & $2.5\times10^{-5}$ \\
\bottomrule
\end{tabular}
\end{table}

For the $3\times2$ lattice, an independent exact-Gauss fPEPS ($n=3$) gives
$E=-32.61$ to $-32.64$, below the DMRG $\chi=512$ value; the DMRG $3\times2$
energy is therefore not converged.

\paragraph{Product-state penalty check.} With all links in the singlet
$\ket5$, the electric terms vanish on a product state, and the penalty reduces
in expectation to $\sum_{v,a}(Q_v^a)^2$. Using
$\sum_a Q_v^aQ_v^a=(3/2)n_v-(3/4)n_v^2$, a singly occupied site contributes
$3/4$, while an empty or doubly occupied (colour-singlet) site contributes
zero. For the \texttt{mixed} state, where every site is singly occupied, the
MPO reproduces $3n_V/4$ for all five geometries ($3.0$, $4.5$, $6.0$, $7.5$,
$6.75$); for the \texttt{double} state, with half the sites doubly occupied and
half empty, it gives zero. The \texttt{mixed} check tests the quadratic terms
and the \texttt{double} check the quartic cancellation, complementing the ED
comparison on a physical state where the penalty vanishes.

\paragraph{Penalty-strength check.} To verify that the accepted $2\times3$ point
is not controlled by the finite penalty, we repeated $g=1$, $m=0.1$ with
$\lambda=100$. At $\chi=448$ the mixed and random states give
$E_0=-32.699007$ and $-32.699013$ with $G^2\lesssim10^{-8}$; at $\chi=512$ the
mixed state gives $E_0=-32.699010$ and $G^2=5.0\times10^{-10}$. The energies
agree with the $\lambda=10$ value $-32.699014$ within $10^{-5}$, and $G^2$
remains below $10^{-8}$. The $\lambda=10$ result is therefore not limited by the
Gauss penalty.

\section{Scalability of the MPS approach}\label{app:scaling}

Table~\ref{tab:scaling} collects the resources that limit the lattice size.
The ED dimension grows as $5^{n_L}4^{n_V}$, so ED is available only on
$2\times2$. For DMRG, the MPS chain length and the MPO bond dimension grow with
the number of links and plaquettes; in addition, the bond dimension $\chi$
required to represent the ground state grows with the entanglement across a cut
through the ladder. With the per-site penalty, the $2\times3$ residuals are
$10^{-11}$--$10^{-14}$. At $\chi=1024$ the $2\times4$ and $2\times5$ ladders
reach $G^2=3.5\times10^{-11}$ and $2.3\times10^{-9}$ and are controlled at the
$10^{-5}$ energy level. The half-filled $3\times3$ pilot has $N_f=9$ (odd).
Since the fermions carry half-integer colour charge and the links integer colour
charge, the summed Gauss law cannot vanish; $G^2=3/4$ is the exact lower bound,
and the DMRG reaches it (0.750 at $\chi=512$ for both $\lambda=10$ and $100$).
With one hole or one doublon ($N_f=8$ or $10$) the residual is
$6.2\times10^{-4}$ at $\chi=512$, $1.1$--$2.2\times10^{-4}$ at $\chi=1024$, and
$1.6$--$2.5\times10^{-5}$ at $\chi=2048$, with the truncation error still
decreasing, so it is limited by $\chi$ rather
than by a $3/4$ floor. The earlier snake run ($G^2=2.25$) lies above this floor
and is not informative about the obstruction. A $3\times2$ pilot with
$L_x=3$ and $N_f=6$ (even) admits an exact-Gauss physical state --- an
independent fPEPS with $n=3$ gives $E=-32.61$ to $-32.64$ with the Gauss law
exact --- so $L_x=3$ is not by itself an obstruction. The DMRG energy at
$\chi=512$ ($-32.5928$, $G^2=6.7\times10^{-11}$ from the random state; the
mixed state, a local minimum, gives $1.9\times10^{-8}$) lies above this
reference, and at $\chi=768$ the penalty-limited state ($E=-32.68$,
$G^2=1.1\times10^{-4}$) is unphysical. The $3\times2$ DMRG point is not a
converged energy benchmark.

\begin{table}[H]
\centering
\footnotesize
\caption{Resource scaling of the MPS/DMRG approach. $n_L$ is the number of
links, $n_V=L_xL_y$ the number of sites, and $N_{\rm MPS}=2n_V+n_L$ the MPS
chain length; $\chi$ is the MPS bond dimension, and the MPO bond dimension is
reported separately. The last column gives the status of the calculations
reported in this paper.}
\label{tab:scaling}
\begin{tabular}{lccccccc}
\toprule
geometry & $n_L$ & $n_V$ & $N_{\rm MPS}$ & ED dimension & MPO bond dim. &
MPS $\chi$ & status \\
\midrule
$2\times2$ & 4 & 4 & 12 & $1.6\times10^{5}$ & 32 & 256 & ED + DMRG \\
$2\times3$ & 7 & 6 & 19 & $3.2\times10^{8}$ & 40 & 512 & DMRG \\
$2\times4$ & 10 & 8 & 26 & $6.4\times10^{11}$ & 40 & 1024 & converged ($10^{-5}$) \\
$2\times5$ & 13 & 10 & 33 & $1.3\times10^{15}$ & 40 & 1024 & converged ($10^{-5}$) \\
$3\times2$ & 7 & 6 & 19 & $3.2\times10^{8}$ & 71 & 768 & even-filling; not converged \\
$3\times3$ & 12 & 9 & 30 & $6.4\times10^{13}$ & 71 & 2048 & half-filling obstructed ($G^2=3/4$) \\
\bottomrule
\end{tabular}
\end{table}

\section{Worked example on the \texorpdfstring{$2\times2$}{2x2} lattice}\label{app:worked}

We spell out the conventions of Section~\ref{sec:model} on the smallest lattice,
$L_x=L_y=2$, with sites
$A=(0,0)$, $B=(1,0)$, $C=(0,1)$, $D=(1,1)$ and links
$h_0=A$--$B$, $h_1=C$--$D$, $v_0=A$--$C$, $v_1=B$--$D$. The plaquette is
$A\to B\to D\to C\to A$, so
$P=\operatorname{Tr}(U_{h_0}U_{v_1}U^\dagger_{h_1}U^\dagger_{v_0})$.

The Gauss generators follow the rule that the left/lower endpoint contributes
$E_L^a$ and the right/upper endpoint contributes $E_R^a$:

\begin{table}[H]
\centering
\footnotesize
\caption{Gauss generators on the $2\times2$ lattice.}
\begin{tabular}{lc}
\toprule
site $v$ & $G_v^a$ \\
\midrule
$A=(0,0)$ & $E_L^a(h_0)+E_L^a(v_0)+Q_A^a$ \\
$B=(1,0)$ & $E_R^a(h_0)+E_L^a(v_1)+Q_B^a$ \\
$C=(0,1)$ & $E_L^a(h_1)+E_R^a(v_0)+Q_C^a$ \\
$D=(1,1)$ & $E_R^a(h_1)+E_R^a(v_1)+Q_D^a$ \\
\bottomrule
\end{tabular}
\end{table}

For the Jordan--Wigner string the site ordering is row-major,
$\sigma(A)=0$, $\sigma(B)=1$, $\sigma(C)=2$, $\sigma(D)=3$; the intermediate
sites and the resulting factors are:

\begin{table}[H]
\centering
\footnotesize
\caption{Jordan--Wigner factors for the four bonds of the $2\times2$ lattice.
The string runs over the sites strictly between $\sigma(v)$ and $\sigma(w)$ in
the row-major ordering.}
\begin{tabular}{lccc}
\toprule
bond & $\sigma(v)\to\sigma(w)$ & intermediate & JW factor \\
\midrule
$h_0:A$--$B$ & $0\to1$ & --- & $1$ \\
$h_1:C$--$D$ & $2\to3$ & --- & $1$ \\
$v_0:A$--$C$ & $0\to2$ & $B$ & $\Pi_B=(-1)^{n_B}$ \\
$v_1:B$--$D$ & $1\to3$ & $C$ & $\Pi_C=(-1)^{n_C}$ \\
\bottomrule
\end{tabular}
\end{table}

The MPS chain is
\[
\underbrace{c_{A,0},\,c_{A,1},\,h_0,\,c_{B,0},\,c_{B,1},\,v_0,\,v_1}_{k=0,\dots,6\ (y=0)},
\qquad
\underbrace{c_{C,0},\,c_{C,1},\,h_1,\,c_{D,0},\,c_{D,1}}_{k=7,\dots,11\ (y=1)},
\]
so $N_{\rm MPS}=2L_xL_y+n_L=8+4=12$. The four link sites have trivial fermion
parity and do not change the JW product.

\paragraph{Basis and action of the Hamiltonian.}
A basis state of this chain is
\[
\ket{m_0,\dots,m_{11}},\qquad
m_k\in\{0,1\}\ \text{(fermion modes)},\quad
m_k\in\{1,\dots,5\}\ \text{(links)},
\]
where $m_k$ is the physical index of the $k$-th chain site, not the site
position $\sigma(v)$. The Hilbert-space dimension is $2^8\cdot5^4=4^4\cdot5^4=160{,}000$, the ED
dimension. The Hamiltonian acts term by term: $E_l^2$ is diagonal on one link
site ($0$ on $\ket{5}$, $3/4$ on $\ket{1}$--$\ket{4}$); the mass term is diagonal
on the two colour modes of a vertex; the hopping term is off-diagonal on the two
fermion modes and the bond link; and the plaquette term is off-diagonal on the
four links $k=2,5,6,9$. Here and below, a ket such as
$\ket{B_0=1,h_0=5}$ lists only the non-default local states; all omitted
fermion modes are empty and all omitted links are $\ket5$. For example, the $h_0$ term alone acts as
\[
-t\left(c^\dagger_{A,0}U_{00}+c^\dagger_{A,1}U_{10}\right)c_{B,0}
\ket{B_0=1,h_0=5}
=t\sqrt2\,\ket{A_0=1,h_0=3}
-t\sqrt2\,\ket{A_1=1,h_0=1},
\]
because the occupied $B_0$ mode fixes $\beta=0$, and
$U_{00}\ket5=-\sqrt2\ket3$, $U_{10}\ket5=+\sqrt2\ket1$. Its Hermitian
conjugate vanishes because $A$ is empty. The full $H_{\rm hop}$ also contains
the $v_1$ channel, which moves the fermion from $B$ to $D$; we isolate the
$h_0$ term here to exhibit the colour contraction and the link update.

On the vertical bond $v_0:A$--$C$ the corresponding term is
\[
\begin{aligned}
&-t\,\Pi_B\left(c^\dagger_{A,0}U_{00}+c^\dagger_{A,1}U_{10}\right)c_{C,0}
\ket{C_0=1,v_0=5}\\
&=t\sqrt2\,\Pi_B\,\ket{A_0=1,v_0=3}
-t\sqrt2\,\Pi_B\,\ket{A_1=1,v_0=1},
\end{aligned}
\]
where $\Pi_B=(-1)^{n_B}$ is the Jordan--Wigner string of the intermediate vertex $B$
(the sites $A$ and $C$ are separated by $B$). For the state written here $B$ is
empty and $\Pi_B=+1$; if $B$ is singly occupied, $\Pi_B=-1$ and both amplitudes
change sign. Its Hermitian conjugate vanishes because $A$ is empty. Horizontal
bonds such as $h_0$ have no such factor because their endpoints are adjacent.
The full $H_{\rm hop}$ also contains the $h_1$ channel, which moves the fermion
from $C$ to $D$.

The plaquette operator is
$P=\operatorname{Tr}(U_{h_0}U_{v_1}U^\dagger_{h_1}U^\dagger_{v_0})$. It is
off-diagonal in the link basis: $P\ket{5,5,5,5}$ contains the component
$4\ket{3}_{h_0}\ket{3}_{v_1}\ket{2}_{h_1}\ket{2}_{v_0}$, so it does not
annihilate the all-singlet configuration; the magnetic term creates and
annihilates rishon pairs. The electric term is instead diagonal:
$E_l^2=0$ on $\ket5$ and $3/4$ on $\ket{1}$--$\ket{4}$.

Finally, with all links in $\ket5$ and one fermion per site,
$\sum_a Q_v^aQ_v^a=3/4$ and the penalty expectation is
$4\cdot3/4=3=3n_V/4$, consistent with the product-state check of
Appendix~\ref{app:params}.

\section*{Data and code availability}
The sparse ED, MPO, and DMRG implementations, the fPEPS cross-check, the driver
scripts, and the per-run JSON files are available from the author on reasonable
request. A public repository with a DOI, the environment specification
(TeNPy 1.1.1, SciPy, quimb, symmray), the command lines used for each table, and
the JSON schema for the per-run diagnostics will accompany the published
version. The numerical values in the tables of this paper can be reproduced from
these files.

\end{document}